\documentclass[12pt,preprint]{aastex}
\usepackage{graphics}
\usepackage{aastexug}  
\usepackage{natbib}

\def\simgt{\lower.5ex\hbox{$\; \buildrel > \over \sim \;$}}
\def\simlt{\lower.5ex\hbox{$\; \buildrel < \over \sim \;$}}
\def\ltsima{$\; \buildrel < \over \sim \;$}
\def\gtsima{$\; \buildrel > \over \sim \;$} 
\def\lsim{\lower.5ex\hbox{\ltsima}} 
\def\gsim{\lower.5ex\hbox{\gtsima}} 
\def\msun{${\rm M_\odot}$} 
 
\begin{document} 
 
\title{Towards a working model for the abundance variations within 
Globular Clusters stars}
 
\author{Paolo Ventura  \altaffilmark{}
\&  Francesca D'Antona \altaffilmark{}
}
 
\altaffilmark
{INAF - Osservatorio  Astronomico  di  Roma, via Frascati
33, 00040 Monteporzio, Italy; dantona, ventura@mporzio.astro.it}

\begin{abstract}  

A popular self--enrichment scenario for the formation of globular clusters 
assumes that the abundance anomalies shown by the stars in many clusters are 
due to a second stage of star formation occurring from the matter lost by the 
winds of massive asymptotic giant branch (AGB) stars. Until today, the
modellizations of the AGB evolution by several different groups failed,
for different reasons, to account for the 
patterns of chemical anomalies. Here we show that our own modelling can
provide a consistent picture if we constrain the three main 
parameters which regulate AGB 
evolution: 1) adopting a high efficiency convection model; 2) adopting 
rates of mass loss with a high dependence on the stellar luminosity; 3) assuming
a very small overshooting below the formal convective regions during the
thermal pulse (TP) phase. The first assumption is needed to obtain an efficient
oxygen depletion in the AGB envelopes, and the second one
is needed to lose the whole stellar envelope within few thermal
pulses, so that the sum of CNO elements does not increase too much,
consistently with the observations. The third assumption is needed
to fully understand the sodium production. We also show that the 
Mg -- Al anticorrelation is explained adopting the higher limit of the NACRE
rates for proton captures by $^{25}$Mg and  $^{26}$Mg, and the models are 
consistent with the recently discovered F-Al correlation.
Problems remain to fully explain the observed Mg isotopes 
ratios. \end{abstract}


\section{Introduction}
\label{sec:intro}
Recent years have witnessed exciting developments both in the observations and 
theoretical modelling of the abundance star to star variations within most of 
the well studied Globular Clusters (GCs). In most GCs\footnote{We leave aside 
the very peculiar clusters like $\omega$Cen, in which the heavy elements spread 
and the HR diagram morphologies clearly show that we are dealing with several 
stellar generations, enriched by the supernova ejecta.} only the light elements 
that are susceptible to abundance changes from proton-capture reactions, such 
as the pp, CN, ON, NeNa, and MgAl cycles, exhibit star-to-star abundance 
variations, far in excess of the modest dispersion seen in halo field stars --
see, e.g., \citet{smi87}, Kraft (1994)\nocite{kra94}, Sneden 
(1999)\nocite{sne99}, \cite{ramirez1, ramirez2}. Observations of 
these abundance spreads at the turnoff and among the subgiant stars 
\citep[e.g.,][]{gratton2001} have shown that these anomalies must be attributed 
indeed to some process of self--enrichment occurring at the first stages of the 
life of the cluster. The massive AGB envelopes are the ideal place to 
manifacture elements through nuclear reactions in which proton captures are 
involved, as, especially for low metallicity, they are subject to hot bottom 
burning (HBB) \citep[e.g.][]{ventura2001, ventura2002}: the convective envelope 
reaches the hydrogen burning shell, and the products of burning are convected 
to the surface of the star and are given back to the intracluster medium by 
means of the stellar wind and the planetary nebula ejection.

Today, the most popular explanation of the observed chemical anomalies, first 
proposed by Cottrell \& DaCosta (1981) and D'Antona, Gratton \& Chieffi 
(1983), and recently\footnote{See, e.g. the Proceedings of the Joint Discussion 
4 of the IAU General Assembly 2003 \citep{dantonadacosta2004}} adapted, is that 
these winds, collected in the central regions of the cluster, initiate a second 
stage of more or less continuous star formation lasting not more than 200Myr. 
This model, which we will now call the {\sl standard self-- enrichment 
scenario}, received {\bf an interesting hint} by the interpretation of  
some peculiar horizontal branch (HB) morphologies in terms of helium 
enrichment of the gas from 
which a fraction of stars were born \citep{dantona2002, dantona2004}. The 
chemical anomalies are correlated with the HB morphology (e.g. Catelan \& 
Freitas Pacheco 1995), and high helium abundances (which are also found in the 
envelopes of the massive AGBs subject to HBB) seem to be the best explanation 
for the HB blue tails and for the peculiar blue main sequences in $\omega$Cen 
\citep{norris2004, piotto2005} and in NGC2808 \citep{dantona2005}. 

In spite of the appealing features of the standard self--enrichment scenario, a 
quantitative reproduction of the observed abundance spreads ---mainly of the 
oxygen vs. sodium and of the aluminum versus magnesium anticorrelations, for 
which abundant data are given in the recent astronomical literature--- is not 
available \citep{denis2003, ventura2004, fenner2004}.  In the recent literature 
indeed it is stated that quantitative considerations reveal a number of serious 
problems and ``the current theoretical yields and chemical evolution models do 
not favour the AGB pollution scenario (with intermediate-mass AGBs) as the 
mechanism responsible for star-to-star abundance variations in globular 
clusters" \citep{yong2005} or that ``the notion that massive AGB stars are the 
origin of the O--Na abundance anticorrelation in GC giants is not 
consistent with the model predictions of this study" \citep{herwig2004}. Most 
of these problems, however, are present in the computations which rest on MLT 
standard stellar models of low convective efficiency, due to which 
the HBB temperatures are not large enough to allow efficient ON cycling. The 
recent models by \cite{fenner2004}, who indeed take care of developping an 
entire chemical evolution model, fail to reproduce the O--Na anticorrelation 
and most of the other chemical anomalies mainly due to this choice of 
convection modelling. 
By tradition, our group adopts the Full Spectrum of Turbulence (FST) model by 
\cite{cm1991, cgm1996}, whose high efficiency allows strong ON cycling.
Recently, \cite{ventura-dantona2005a} have shown in detail that the 
modelization of the nuclear yields is enormously dependent on the efficiency of 
the adopted convection model and, indirectly, on the efficiency of mass loss. 
Some problems of other modellers, e.g. the high increase in the CNO total 
abundances, which {\it is not} found in the observations \citep{ivans1999, 
briley2002, briley2004, cohen-melendez2005, cohen2005}, is in fact due to the 
high number of third dredge up (TDU) episodes, due to the comparatively low 
luminosity of the models with respect to the luminosity of the FST models.
In spite of this improvement, our most recent models 
\citep{ventura-dantona2005b} reproduce in a satisfactory way the O--Na 
anticorrelation only for a limited range of masses (3.5 - 4.5\msun), 
and are not consistent with the Mg-Al anticorrelation. 

In this paper, we show that we can  modify our models in order to solve these 
problems, and thus we are closer to the complete solution of  this appealing 
conundrum. The guide-line of this study was to build AGB stellar models whose 
ejecta are consistent with the chemistry of the most contaminated stars within 
GCs, i.e.: a) strong enhancement of aluminum and fluorine; b) a positive yield 
of sodium; c) a large depletion of oxygen; d) a (C+N+O) sum increased by no 
more than a factor of $\sim 2$.   

We show that the correlation of sodium with aluminum, and the anticorrelations 
of sodium versus oxygen and fluorine, can all be reproduced by assuming a 
modest extra-mixing from the base of the convective envelope in addition to a 
very efficient convection model.

\section{Modelling the AGB phase}
\noindent
We focus our attention on a typical massive AGB model, of mass M=5\msun;
the adopted chemistry is (Z,Y)=(0.001,0.24). The complete description of 
the physics of the models is given in 
\cite{ventura-dantona2005a, ventura-dantona2005b}. The detailed study of the 
production of elements during the AGB evolution requires computation of 
complete stellar models including nuclear processing by HBB, coupled with 
non--instantaneous mixing.

We adopt a network of 30 isotopes from hydrogen to silicon. The standard cross 
sections adopted in the code are from \cite{nacre}. The abundances 
we adopt in the computation are solar-- scaled, due to the present 
limitation of opacities and equation of state in our code. 
Actually, the $\alpha$ elements in popII stars are enhanced by
$\sim 0.2 -- 0.4$ dex (e.g. Gratton, Sneden \& Carretta 2004), and
the initial value of CNO abundances affect the detailed nucleosynthesis.
However we have made test models concerning the important oxygen
abundance bahaviour, showing that the logarithmic decrease of
oxygen abundance is roughly the same. Consequently, if we had started, 
e.g., from an oxygen overabundance of +0.3dex, the final abundance 
would have been 0.3dex larger than that found in our model. 

Convection is modelled according to the Full Spectrum of Turbulence (FST) 
treatment by \cite{cgm1996}, which provides at the bottom of the convective
envelopes temperature larger than the corresponding MLT models with
$l/H_p=2$ \citep{ventura-dantona2005a}.

The standard mass loss law adopted in our models \citep[e.g.][]{mazzitelli1999} 
is from \cite{blocker}, who modifies the Reimer's formula in order to simulate 
the strong mass loss suffered by these stars as they climb along the AGB.
The $\eta$ parameter in this formulation is set to 0.02. 

\section{Comparison of models with observations}

The strong depletion of oxygen needed to match the observed abundances of the 
bulk of the stars showing chemical anomalies in GCs  ($\sim -0.5$ dex) requires 
very high temperatures at the base of the external envelope, that can be hardly 
achieved in the framework of the MLT modelling of convection; the recent 
observations of giant stars in M4 \citep{smith2005}, indicating a strong 
fluorine depletion ($\sim -0.8$ dex), also point towards the same direction. 
The problem most difficult to overcome is that models depleting oxygen destroy 
sodium as well \citep{denis-weiss2001}. When an efficient TDU is invoked,  the 
sodium abundance can increase by even a factor of $\sim 100$, that is at odds 
with the highest observed value in the M13 giants ($\sim +0.6$ dex); besides, 
in these same models  the (C+N+O) sum would be much larger than observed.
As discussed, the models by \cite{ventura-dantona2005a} 
reproduced the oxygen depletion and sodium enhancement only
for a small range of masses. In addition,
the range of $^{27}Al$ enhancement was a 
major problem of these models.

We investigate now the effects that a small amount of extra-mixing from 
the base of the convective envelope might have on the chemistry of the 
ejecta, by assuming that convective velocities
decay exponentially inwards beyond the formal border fixed by the
Schwartzschild criterion. The difference between the present model
and the model of the same mass presented in Ventura \& D'Antona (2005b)
is that the parameter $\zeta$ connected with the extra-mixing is
not set to zero, but is allowed to vary between $0\leq \zeta \leq 0.02$,
where $\zeta=0.02$ is the value used to simulate overshooting 
from the border of the convective core of intermediate mass stars
during the phase of H-burning, that was calibrated in order to reproduce
the observed main sequences of open clusters (Ventura et al. 1998).

Figs.~\ref{ejecta1} and \ref{ejecta2} show the evolution (in terms 
of variation with mass, that is obviously decreasing with time) 
of the surface abundances of some key-elements of 
5\msun models calculated with different $\zeta$'s. 
First, we note that large depletion of oxygen and fluorine can be
easily achieved when the FST treatment is used for convection modelling.
We also confirm, (left panel of 
Fig.~\ref{ejecta1}, solid line) that sodium cannot be produced if a straight
Schwartzschild criterion is used to locate the base of the surface
convection. Conversely, if some extra-mixing is assumed, 
the surface sodium increases following each TP, due to the 
dredge-up of $^{22}Ne$, later converted to $^{23}Na$; this
holds particularly during the latest evolutionary stages, when 
the temperature at the base of the surface convective region is 
decreased by mass loss.
Another bonus of these models is that the small total number of TPs prevents 
from obtaining a final sodium abundance largely exceeding the initial value.

Fig.~\ref{ejecta1} (right panel) shows that the larger is
$\zeta$ the larger is the oxygen content of the ejecta, as this latter 
element can also be dredged-up following each TP if the base of the 
convective envelope penetrates deeply into the ashes of the former 
$3\alpha$ shell. We find that a value $\zeta \sim 0.0015$ leads
to ejecta whose chemical composition is both sodium rich ($[Na/H]=+0.4$)
and oxygen poor ($[O/H]=-0.55$), and can thus take account of the 
observed anticorrelation sodium versus oxygen. 
The use of $\zeta=0.0015$ leads to a dredge-up
parameter $\lambda$ gradually increasing along the AGB evolution,
that reaches a maximum value of $\lambda=0.55$ at the latest TPs. {\bf 
This, combined with the small number of TPs in which TDU is achieved
($\sim 10$), makes the mass $M_{proc}$ of the processed material to be 
$M_{proc} < 5\times 10^{-3}$; we therefore expect a negligible effect
on the s-process elements abundances.}

The left panel of Fig.~\ref{ejecta2} also shows that a
strong depletion of fluorine is achieved in all cases; the
mass expelled by the $\zeta=0.0015$ model has 
$[F/H] \sim -1.2$, a result that is in agreement with the
analysis by Smith et al. (2005) regarding the fluorine abundance
variations in red giants of M4 (see their Fig.4).
The central panel shows that when using the higher limit of the NACRE 
rates for $^{25}Mg$ and $^{26}Mg$ proton capture reactions 
it is also possible to achieve a strong $^{27}Al$ production,
($[^{27}Al/H] = +0.7$) independently of $\zeta$;
this can not be obtained when we adopt the recommended NACRE cross sections 
(long-dashed line).
Finally, the right panel of Fig.~\ref{ejecta2} shows the evolution
of the total (C+N+O) abundance, with respect to the initial value.
We see that the model calculated with $\zeta=0.0015$ achieves a
maximum final abundance increased by a factor of $\sim 3$, and the
average increase is $\sim 1.5$, that is still consistent with the
observations. 
{\bf Concerning the magnesium isotopic ratios, the ejecta
of our model have 
$^{25}Mg/^{24}Mg \sim 7$ and $^{26}Mg/^{24}Mg \sim 2$, 
both sensibly greater than the values ($\sim 1$) observed 
by Yong et al.(2005) in the most polluted stars of NGC6752: 
we recall, however, that the theoretical estimate of relative 
abundances of the magnesium isotopes is affected by the 
uncertainties related to the cross section of the reactions 
of proton and $\alpha$ captures by the heavy magnesium isotopes, 
and the $\alpha$ captures by $^{22}Ne$ nuclei.}
    
\section{Conclusions}
In this work we present results of detailed calculations focused on
the AGB phase of a typical intermediate mass star (M$=5$\msun) of
metallicity $Z=0.001$.

Our main finding is that when the FST model is used to deal with convective 
modelling, the use of a modest amount of extra-mixing at the base of the outer 
convective envelope leads to ejecta whose chemical composition reproduces 
simultaneously most of the chemical anomalies observed in GCs stars. 

In particular, we find that:

\begin{enumerate}

\item{Oxygen and fluorine are strongly depleted: we find
$[O/H]=-0.55$ and $[F/H]=-1.2$.}

\item{When the high NACRE limit for the rates of the proton capture
reactions by the heavy magnesium isotopes are used, aluminum
can be easily manifactured: $[^{27}Al/H]=+0.7$.}

\item{A modest amount of extra-mixing at the base of the convective
envelope ($\zeta=0.0015$, to be compared with the value $\zeta=0.02$
that must be used in our models to increase the convective core extension during 
H-burning, to fit the observed open cluster main-sequences) triggers
some $^{22}Ne$ to be dredged-up following each TP; when the pulse is
extinguished, this fresh $^{22}Ne$ is converted in sodium, that in
the latest evolutionary stages, when the temperature at the base of the 
convective envelope is reduced by mass loss, can survive to proton 
fusion. The overall sodium content of the ejecta is $[^{23}Na/H]=+0.4$}

\item{The total (C+N+O) abundance, due both to the small number of TPS
and to the modest efficiency of the TDU, is increase by only
a factor of $\sim 1.5$, that is also consistent with the observations.}

\end{enumerate}

The present results give more robustness to the current interpretation 
that the chemical anomalies among GC stars result from star formation in 
the intracluster medium polluted by the ejecta of an early generation of 
massive AGBs.

\begin{figure*} 
\caption{Variation with mass of the 
surface abundances of sodium (Left) and oxygen (Right) within models of 5\msun 
calculated with different assumptions regarding the extra-mixing region from 
the base of the surface convective mantle. On the vertical axis we report for
each chemical the quantity $[X]=\log(X/X_i)$, where $X_i$ is the initial
abundance.} 
\label{ejecta1} 
\end{figure*}

\begin{figure*}
\caption{Variation with mass of the surface abundances of
fluorine (Left), aluminum (Centre) and of the ratio
of the total (C+N+O) abundance compared to the initial
value (Right). The long-dashed track in the central panel
gives the predicted aluminum abundance when the recommended
NACRE cross-section for the proton capture reactions by the
heavy magnesium isotopes are used.}
   \label{ejecta2}
\end{figure*}

\end{document}